\documentclass[%
aip,apl,
numerical,
amsmath,amssymb,
reprint,%
]{revtex4-2}

\usepackage{graphicx}
\usepackage{dcolumn}
\usepackage{bm}

\usepackage[utf8]{inputenc}
\usepackage[T1]{fontenc}
\usepackage{mathptmx}
\graphicspath{{Figures/}{./}} 

\begin{document}


\title{A multiphysics model for high frequency optomechanical sensors optically actuated and detected in the oscillating mode} 



\author{S. Sbarra}
\altaffiliation{}
\affiliation{Matériaux et Phénomènes Quantiques, Université de Paris, CNRS, UMR 7162, 10 rue Alice Domon et Léonie Duquet, Paris 75013, France}

\author{P. E. Allain}
\altaffiliation{}
\affiliation{Matériaux et Phénomènes Quantiques, Université de Paris, CNRS, UMR 7162, 10 rue Alice Domon et Léonie Duquet, Paris 75013, France}

\author{A. Lemaître}
\altaffiliation{}
\affiliation{Centre de Nanosciences et de Nanotechnologies, CNRS, UMR 9001, Université Paris-Saclay, Palaiseau 91120, France}

\author{I. Favero}
\email[]{ivan.favero@u-paris.fr}
\altaffiliation{}
\affiliation{Matériaux et Phénomènes Quantiques, Université de Paris, CNRS, UMR 7162, 10 rue Alice Domon et Léonie Duquet, Paris 75013, France}

\date{\today}

\begin{abstract}
Optomechanical systems combine extreme sensitivity and bandwidth in the control of mechanical motion, of interest for various applications. Integrated on a chip, actuated and detected all-optically by a single laser, they could disrupt sensing technologies. We introduce here a multiphysics model that describes their operation in the oscillating mode, under sinusoidal modulation of the laser, when both photothermal forces and radiation pressure are present, and when nonlinear absorption occurs in the device. The model is validated by systematic experiments on ultra-high frequency optomechanical disk resonators and leads to a quantitative assessment of the amplitude and phase of the demodulated output signal, which carries the sensing information.
\end{abstract}


\maketitle 

\noindent The small dimensions of mechanical micro- and nano-resonators induces a large responsivity to external perturbations, making these systems ideal for sensing purposes \cite{Hanay2012}$^{,} $\cite{Sage2018}. Actuation of the mechanical system is necessary to increase the vibration amplitude and improve its sensing performances \cite{Sansa2016a}$^{,} $\cite{Ekinci2004b}.
Among multiple actuation mechanisms, optical driving of mechanical resonators enables broadband actuation up to the GHz mechanical frequency range. At the same time, optical techniques permit ultrasensitive, eventually quantum-limited, detection of motion. For these reasons, several optomechanical devices \cite{Favero2009}$^,$ \cite{Aspelmeyer2014a} have been pushed forward for magnetic field \cite{Forstner2012}, mass \cite{Liu2013a}$^,$ \cite{Yu2016}$^,$ \cite{Sansa2020a} or atomic force sensing \cite{Chae2017}$^,$\cite{Allain2020}. Driving and detecting the mechanical sensor in an all-optical way, with a single laser source,
offers an obvious advantage of simplicity, well suited for integration. The oscillating sensing mode, where the mechanical system is sinusoidally forced, is then obtained under coherent modulation of the laser, while the output light is demodulated.

Early experiments in optomechanics, while not aiming at sensing, did implement such modulation/demodulation approach in order to characterize the dynamical response of the system under study \cite{Schliesser2006}$^,$ \cite{Metzger2008}. In$\;$ \cite{Metzger2008}, the effect of photothermal forces, where photons are absorbed and thermally distort the mechanical system, was considered within a delayed force model. The latter efficiently depicted the behaviour of employed cantilevers of mechanical frequency $\omega_m$=2$\pi$ $\times$10kHz, but was inadapted for high frequency devices operating in the good cavity limit $\omega_m \gtrsim\kappa$ with $\kappa$ the optical cavity decay rate \cite{Restrepo2011}. In contrast, the canonical optomechanical radiation-pressure model \cite{Aspelmeyer2014a} correctly operates at arbitrary high mechanical frequency, and modulation/demodulation experiments in this regime have been popularized as being the optomechanical analogue of electromagnetically induced transparency\cite{Weis2011}$^,$\cite{Safavi-Naeini2011}. Unfortunately, the latter model neglects photothermal interactions, which are often sizable at room temperature and of concrete importance for operational optomechanical sensors \cite{Belacel2017}$^,$\cite{Zhu2016}$^,$\cite{Guha2020}. In a recent paper\cite{Guha2017}, a model was introduced that solved for that discrepancy by writing three coupled equations for the cavity mode, the mechanical and thermal degrees of freedom of a resonator, allowing quantitative modelling of dynamical backaction effects at ultra-high frequency with significant photothermal interactions. Here we explore the modulation/demodulation regime associated to this latter model. We derive compact analytical expressions for both quadratures of the demodulated signal, including in a regime where nonlinear absorption is present, giving rise to a nonlinear component of the optical force. We confront these expressions to systematic experiments on optomechanical disk resonators with mechanical modes in the ultrahigh frequency range, varying the modulation frequency, the optical operating conditions such as detuning and power, and the investigated mechanical modes. The precision of the tested model prepares the ground for the calibrated use of integrated optomechanical sensors optically operated at room temperature, a required step for concrete applications.

\begin{figure}[!h]
	\includegraphics[width=8.5cm]{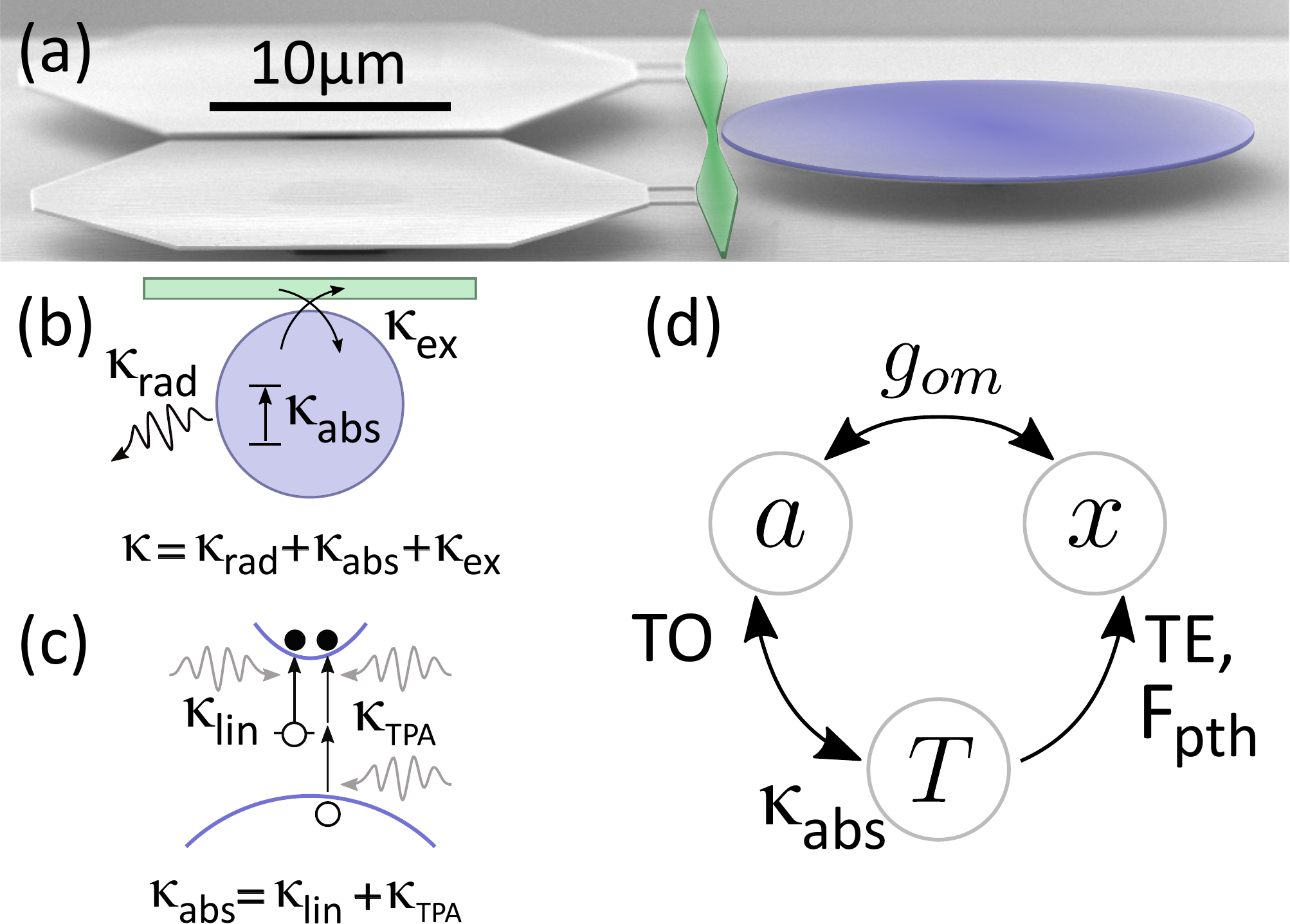}
	\caption{(a) Electron micrograph: GaAs optomechanical disk (blue) in the vicinity of its coupling waveguide (green), whose extremities are tapered for efficient light injection and collection. The guide is supported by two hexagonal holding pads (left), which play no optical nor mechanical role. (b) Three contributions to the optical cavity loss rate ($\kappa$): the radiative losses ($\kappa_\mathrm{rad}$), coupling rate to the waveguide ($\kappa_\mathrm{ex}$) and absorption losses ($\kappa_\mathrm{abs}$). (c) Linear absorption ($\kappa_\mathrm{lin}$) involves single telecom photon processes, while two-photon absorption TPA ($\kappa_\mathrm{TPA}$) involves pairs of photons. (d) Optical, mechanical and thermal degrees of freedom in interaction (see text).}
	\label{fig:Figure1}
\end{figure}

A electron micrograph of the optomechanical system under investigation is shown in Fig.~\ref{fig:Figure1}~(a). 
It consists of a Gallium Arsenide (GaAs) disk patterned on an GaAs(200~nm)/Al$_\mathrm{0.8}$Ga$_\mathrm{0.2}$As(1800~nm)/GaAs(substrate) epitaxial wafer using e-beam lithography and inductively coupled plasma etching. 
Hydrofluoric acid under-etching is employed to selectively remove the AlGaAs and shape the disk pedestal.
This structure supports optical whispering gallery modes (WGMs) that can be excited via an integrated suspended waveguide at a rate $\kappa_\mathrm{ex}$ (Fig.~\ref{fig:Figure1}~(b)). Radiative contributions to the WGM cavity losses (bending and scattering losses) are grouped under the rate $\kappa_\mathrm{rad}$. Intracavity photons are absorbed at a rate $\kappa_\mathrm{abs}$. As depicted in Fig.~\ref{fig:Figure1}~(c), a single telecom (sub-bandgap) photon can be absorbed in a transition involving a mid-gap state \cite{Parrain2015}$^,$\cite{Guha2017} ($\kappa_\mathrm{lin}$), while a pair can be directly absorbed by two-photon absorption ($\kappa_\mathrm{TPA}$), such that $\kappa_\mathrm{abs}=\kappa_\mathrm{lin}+\kappa_\mathrm{TPA}$. Both effects, linear and nonlinear in the circulating power, are responsible for heating up the resonator.
For a $\Delta T$ temperature increase of the disk, the local stress induced by thermal expansion is given by \cite{Favero2020}:
\begin{equation}
	\sigma_{ij}^{th}=C_{ijkl}\beta_\mathrm{th} \delta_{kl}\Delta T
	\label{eq:thermal stress}
\end{equation}
with $C_{ijkl}$ the stiffness tensor and $\beta_\text{th}$ the thermal expansion coefficient of the material. 
Each mechanical mode of the resonator is impacted its own way by this thermal stress. In a lumped element model associated to a given mechanical mode, the effective mass on a spring is subjected to a photothermal force $F_\mathrm{pth}$, whose amplitude is given by\cite{Favero2020}:
\begin{equation}
	F_\mathrm{pth}=\int_V{dV \sigma_{ij}^{th}S_{ij}}=\alpha\times\Delta T
	\label{eq:photothermal force}
\end{equation}
where $S_{ij}$ is the strain field of the considered mechanical mode.
Another consequence of the temperature increase in the disk is the red-shift of optical and mechanical resonances.
The first is a consequence of the thermo-optic effect (TO) while the second is related to the thermo-elastic (TE) softening of the material at high temperature. When combined with the canonical optomechanical coupling between the motion $x$ and the optical cavity field $a$, these various thermal effects give rise to a close set of interactions between optical, mechanical and thermal degrees of freedom (Fig. ~\ref{fig:Figure1}~(d)), governed by three coupled equations: 
\begin{eqnarray}
	\dot{a}=-\frac{\kappa}{2} a+ i \left( {\Delta}+g_\mathrm{om}x+\frac{\omega_\mathrm{cav}}{n}\frac{dn}{dT}\Delta T \right) a +\sqrt{\kappa_\mathrm{ex}} a_\mathrm{in} \nonumber,
	\\
	m_\mathrm{eff} \ddot{x}+m_\mathrm{eff} \Gamma_\mathrm{m}\dot{x}+ m_\mathrm{eff}\omega_m^2x= F_\mathrm{pth}+F_\mathrm{opt} \nonumber,
	\\
	\dot{\Delta T}=-\frac{1}{\tau_\mathrm{th}} \left(\Delta T- R_\mathrm{th} \kappa_\mathrm{abs} \hbar \omega_\mathrm{L} |a|^2\right),
	\label{eq:coupled_equations}
\end{eqnarray}
with $\Delta=\omega_\mathrm{L}-\omega_\mathrm{cav}$ the laser-cavity detuning, $g_\mathrm{om}=-\partial \omega_\mathrm{cav}/\partial x$ the optomechanical frequency-pull parameter, $n$ the refractive index, $dn/dT$ the thermo-optic coefficient. Optical fields are written in the rotating frame.
$|a|^2$ is normalized to the number of photons in the cavity and $a_\mathrm{in}$ such that $\hbar \omega_l|a_\mathrm{in}|^2$ is the input power in the waveguide. $m_\mathrm{eff}$, $\Gamma_\mathrm{m}$ and $\omega_\mathrm{m}$ are the mechanical resonator's effective mass, damping rate and (temperature-dependent through TE) resonant frequency. Forces acting on the mass include a photothermal ($F_\mathrm{pth}$) and a radiation pressure and electrostrictive ($F_\mathrm{opt}$) contribution.
The latter is given by as $F_\mathrm{opt}=\hbar g_\mathrm{om} |a|^2$, where $g_\mathrm{om}$ is calculated numerically considering both the geometrical and photo-elastic coupling \cite{Baker2014b}.
$R_\mathrm{th}$ and $\tau_\mathrm{th}$ are the thermal resistance and relaxation time of the resonator.
This model serves as a starting point to describe modulation/demodulation experiments of interest for sensing in the oscillating mode.

\begin{figure}[!h]
	\includegraphics[width=8.5cm]{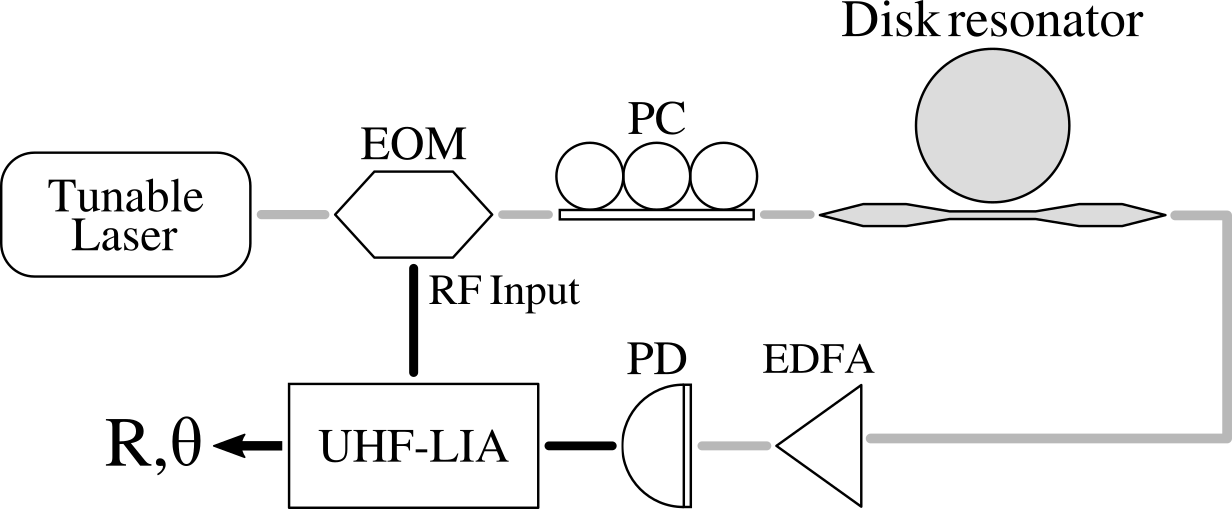}
	\caption{Experimental setup of the all-optical actuation/detection of mechanical motion. The amplitude of the laser power is modulated by an electro-optic modulator (EOM). The polarization is selected with a polarization controller (PC). Light is injected into the integrated waveguide evanescently coupled to the disk resonator, and then collected and amplified by an Erbium-doped Amplifier (EDFA). The signal is converted into current by a high-bandwidth photo-detector (PD) and finally sent to a Ultra-High Frequency Lock-in Amplifier (UHF-LIA), where it is mixed with the reference signal.}
	\label{fig:Figure2}
\end{figure}

Fig.~\ref{fig:Figure2} shows the experimental set-up employed to perform optical actuation and detection of the mechanical device previously described. The light of a tunable telecom laser is amplitude-modulated by a Mach-Zehnder electro-optic modulator (EOM), generating two side-bands in the input field\cite{RogersIII2010a}: $a_\mathrm{in}(t)=\bar{a}_\mathrm{in}\left(1+\beta/2 e^{+i\Omega t}+\beta/2 e^{-i\Omega t}\right)$, where $\Omega$ and $\beta$ are the modulation angular frequency and depth.
Two micro-lensed fibers provide injection into and collection from the waveguide coupled to the micro-disk, where a TE or TM WGM is excited depending on the polarization controller (PC) selection. The intracavity field response to the modulation is \cite{Weis2011}: $a(t)=\bar{a}+\delta a(t)$ with $\delta a(t)=A^-e^{-i\Omega t}+A^+e^{+i\Omega t}$, while we write the displacement and temperature increase: $x(t)=\bar{x}+\delta x(t)$, $\Delta T(t)=\overline{\Delta T}+\delta T(t)$ with $\delta x(t)=Xe^{-i\Omega t}+X^*e^{+i\Omega t}$ and $\delta T(t)=\Delta T_c e^{-i\Omega t}+\Delta T_c^* e^{+i\Omega t}$. The steady-state value of the fields are given by: $\bar{a}=\sqrt{\kappa_\mathrm{ex}}\bar{a}_\mathrm{in}/(k/2-i\bar{\Delta})$, $\bar{x}=(\bar{F}_\mathrm{opt}+\bar{F}_\mathrm{pth})/(m_\mathrm{eff}\omega_\mathrm{m}^2) $, $\overline{\Delta T}=R_\mathrm{th}(\kappa_\mathrm{lin}+\bar{\kappa}_\mathrm{TPA})\hbar\omega_\mathrm{L} |\bar{a}|^2$, with $\bar{\Delta}=\Delta +g_\mathrm{om}\bar{x}+\omega_\mathrm{cav}/n \times dn/dT\times \overline{\Delta T}$ the detuning modified by the optomechanical coupling and the thermo-optic effect.
The output optical signal is first amplified by an Erbium-doped fiber amplifier (EDFA) and converted into an electrical signal by a photo-detector (PD), which is then fed into an Ultra-High-Frequency (UHF) Lock-in Amplifier (LIA).
This latter signal is proportional to the modulus squared of the output light field:
\begin{eqnarray}
	|a_\mathrm{out}(t)|^2=\Big| \bar{a}_\mathrm{in}-\sqrt{k_\mathrm{ex}} \bar{a} +\left(\frac{\beta}{2} \bar{a}_\mathrm{in}-\sqrt{k_\mathrm{ex}} A^- \right) e^{-i\Omega t} \nonumber \\ 
	+\left(\frac{\beta}{2} \bar{a}_\mathrm{in}-\sqrt{k_\mathrm{ex}} A^+ \right) e^{+i\Omega t} \Big|^2
\end{eqnarray} 
which comprises a DC term and two additional components oscillating at $\Omega$ and $2\Omega$. The LIA demodulates this signal at $\Omega$ and decomposes it into an in-phase ($I$) and quadrature ($Q$) component:
\begin{eqnarray}
	I=Re\{(\bar{a}_\mathrm{in}^*-\sqrt{\kappa_\mathrm{ex}}\bar{a})(\bar{a}_\mathrm{in}\beta-\sqrt{\kappa_\mathrm{ex}}(A^-+A^+))\}\nonumber\\
	Q=Im\{(\bar{a}_\mathrm{in}^*-\sqrt{\kappa_\mathrm{ex}}\bar{a})(\sqrt{\kappa_\mathrm{ex}}(A^--A^+))\}
\end{eqnarray}
or into an amplitude ($R=\sqrt{I^2+Q^2}$) and phase ($\theta=\arctan \left(Q/I\right)$). Injecting the field ansatz into the governing equations, we find:
\begin{eqnarray}
	\label{eq:Aplus}
	A^+(\Omega)&=&\frac{i}{2}\frac{2\phi (\Omega+\bar{\Delta} -i \kappa/2) + \left( \phi+\phi^*\right)(\zeta_\mathrm{opt}^*+\zeta_\mathrm{th}^*)} {\bar{\Delta}^2+\bar{\Delta}(\zeta_\mathrm{opt}^*+\zeta_\mathrm{th}^*)-(\Omega-i\kappa/2)^2}	
\end{eqnarray}
\begin{eqnarray}
	\label{eq:Amin}
	A^-(\Omega)&=&\frac{i}{2}\frac{2 \phi (-\Omega+\bar{\Delta} -i \kappa/2) + \left( \phi+\phi^*\right)(\zeta_\mathrm{opt}+\zeta_\mathrm{th})} {\bar{\Delta}^2+\bar{\Delta}(\zeta_\mathrm{opt}+\zeta_\mathrm{th})-(\Omega+i\kappa/2)^2}
\end{eqnarray}
with $\phi=\sqrt{\kappa_\mathrm{ex}} \bar{a}_\mathrm{in} \beta/2$,  $\zeta_{opt}=2 \bar{a}^2\hbar g_\mathrm{om}^2 \chi(\Omega)$, $\zeta_{th}=2 \bar{a}^2(\omega_\mathrm{cav}/n\times dn/dT+\alpha g_\mathrm{om} \chi(\Omega)) R_\mathrm{th}\hbar\omega_\mathrm{L} (\kappa_\mathrm{lin}+2\bar{\kappa}_\mathrm{TPA})(1+i\Omega \tau_\mathrm{th})^{-1}$, $\phi^*$, $\zeta_\mathrm{opt}^*$ and $\zeta_\mathrm{pth}^*$ their complex conjugates, and $\chi(\Omega)=\left[m_\mathrm{eff}(\omega_\mathrm{m}^2-\Omega^2-i\Omega \Gamma_\mathrm{m})\right]^{-1}$ the mechanical susceptibility.
When thermal effects are switched off, and under the approximation of a single sideband in the input field, Eqs.~\ref{eq:Aplus} and ~\ref{eq:Amin} lead back to the results established in the context of optomechanically induced transparency \cite{Weis2011}. Expressions for $\Delta T_c$ and $X$ are given in the supplementary material.
\begin{figure}[!h]
	\includegraphics[width=8.5cm]{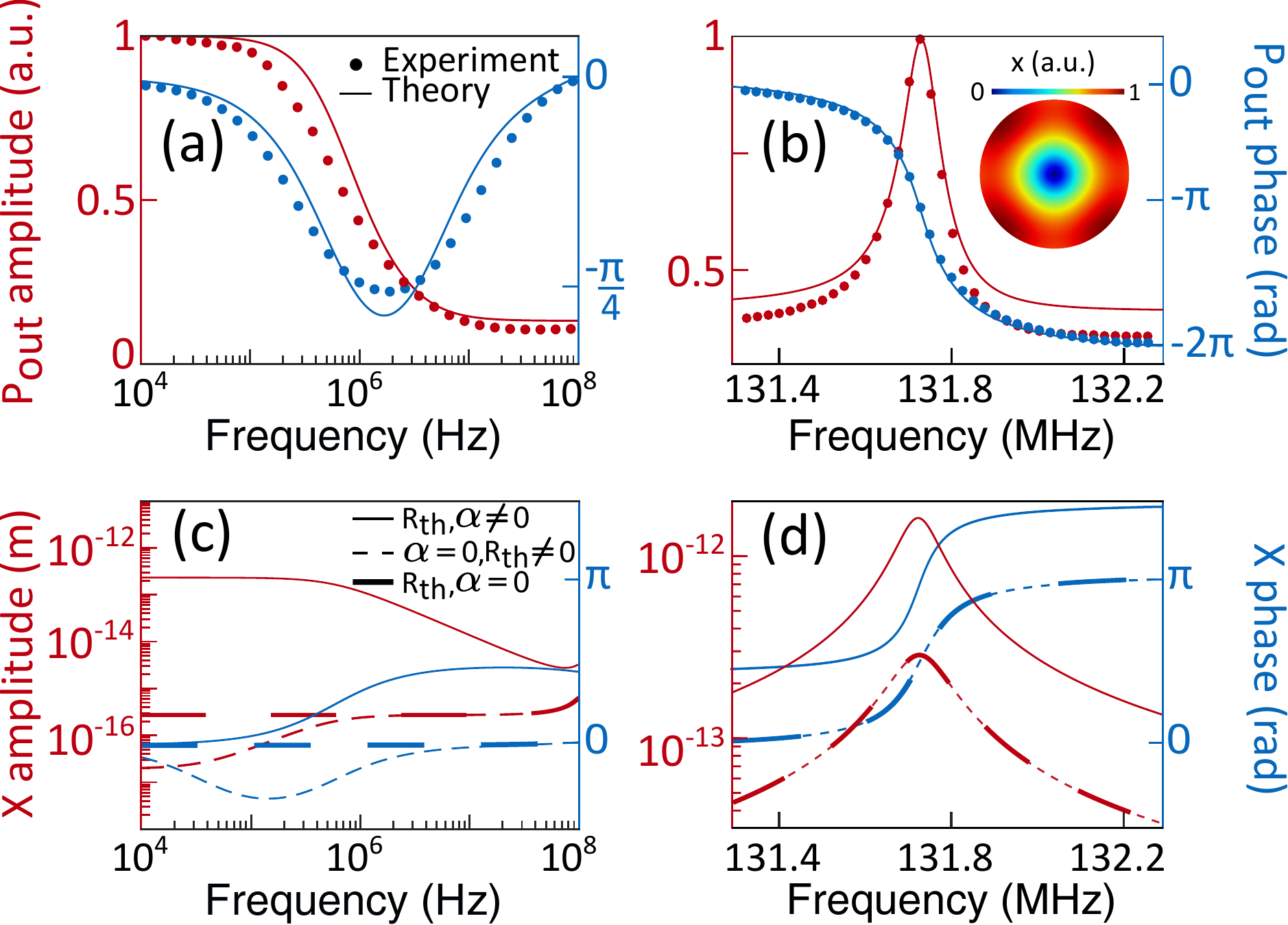}
	\caption{Amplitude and (unwrapped) phase response function of the demodulated optical signal when the modulation frequency is swept between $10$~kHz and $100$~MHz (a) and around the mechanical resonance frequency of RBM1 (b). The mode profile of RBM1 is in inset. The amplitude is normalized by its maximum value over the frequency span. (c),(d): X deduced from the model. The solid line represents the results obtained with the full model when photothemal forces and heating are present ($\alpha \neq 0$, $R_\mathrm{th} \neq 0$), while the thin and thick dashed lines represents respectively the case where photothermal forces ($\alpha = 0$) and heating of the resonator ($R_\mathrm{th}, \alpha = 0$) are switched off.}
	\label{fig:Figure3}
\end{figure}

With the model now in hands, we start by presenting results of modulation/demodulation both at low frequency (10~kHz-100~MHz) and at high frequency, close to the first order Radial Breathing Mode (RBM1), whose resonant frequency is located at $132$~MHz for the present disk (11~$\mu$m radius, 200~nm thickness). An optical power of $200$~$\mu$W is injected in the integrated waveguide and the laser wavelength is tuned to the blue flank of the optical WGM resonance ($\Delta>0$). When sweeping the modulation frequency from $10$ kHz to $100$ MHz, a dip in the phase and a decrease in the amplitude appear in the demodulated signal (Fig.~\ref{fig:Figure3}~(a)), as consequence of a thermal phase lag. Indeed, as apparent in the first line of Eq.~\ref{eq:coupled_equations}, the thermo-optic effect, just as the canonical optomechanical coupling, modifies the amplitude and the phase of the cavity optical field. Being a consequence of photon absorption, the former is filtered by the thermal response of the device (in the microsecond range), and hence distinguishable from the latter. Much larger amplitude and phase shifts in the demodulated signal are however found closer to the resonant frequency of the RBM1 (Fig.~\ref{fig:Figure3}~(b)), whose mode profile is shown in the inset. The photothermal force DC amplitude is three orders of magnitude larger than radiation pressure ($\alpha R_\mathrm{th} \omega_\mathrm{cav} \bar{\kappa}_\mathrm{abs}/g_\mathrm{om}=3\times 10^{3}$), which contributes to an efficient driving of motion even at the frequency of RBM1. The experimental results of Fig.~\ref{fig:Figure3}~(a) and (b) are well reproduced by the model introduced above (Eqs.~\ref{eq:Aplus},\ref{eq:Amin}) (solid line), when using the parameters listed in Table~\ref{Table1}. 
The vast majority of these parameters have been independently measured or calculated with finite element method (FEM), while $\tau_\mathrm{th}$ has been obtained from the fit of the low frequency region ($\leq$100~MHz) (Fig.~\ref{fig:Figure3}~(a)) and found to be consistent with the FEM value. 
The amplitude of the absorptive effects, parametrized by $\kappa_\mathrm{lin}$, $\kappa_\mathrm{TPA}$ and $R_\mathrm{th}$,  was obtained by fitting the thermo-optic shift and distortion of the WGM resonance \cite{Parrain2015} (see supplementary material). The nonlinear absorption rate is proportional to the TPA coefficient $\beta_\mathrm{TPA}$\cite{Johnson2006}:
\begin{equation}
	\kappa_\mathrm{TPA}=\frac{\Gamma_\mathrm{TPA} \beta_\mathrm{TPA} c^2}{V_\mathrm{TPA} n_g^2}\hbar \omega_\mathrm{L} |a|^2
\end{equation} 
with $\Gamma_\mathrm{TPA}$ and $V_\mathrm{TPA}$ the TPA confinement factor and volume, $c$ the speed of light and $n_\mathrm{g}$ the group index. With all parameters of Table~\ref{Table1} fixed this way, the fit of the response in the high frequency region is obtained with no additional adjustable parameter (Fig.~\ref{fig:Figure3}~(b)). 
\begin{table}[!h]
	\caption{\label{Table1} Model Parameters}
	\begin{ruledtabular}
		\begin{tabular}{lccc}
			Parameter&Value&Units&Source\\
			\hline
			$\omega_\mathrm{cav}$ & $2\pi \times 1.93\times 10^{14}$ & Hz & measured \\
			$\kappa_\mathrm{rad}$ & $10.2$ & GHz& measured\\
			$\kappa_\mathrm{ext}$ & $8.00$ & GHz& measured\\
			$\kappa_\mathrm{lin}$ & $0.79$ & GHz& measured\\
			$\beta_\mathrm{TPA}$ & 30 & cm GW$^{-1}$& fit and ref. \cite{Kleinman1973}$^,$ \cite{Krishnamurthy2011}$^,$ \cite{Hurlbut2006}\\
			$\Gamma_\mathrm{TPA}$ & 0.9994 &-& FEM\\
			$V_\mathrm{TPA}$ & $2.42 \times 10^{-17}$ & m$^3$& FEM\\
			$n_\mathrm{g}$ & 3.53 & -& ref. \cite{Skauli2003}\\
			$P_\mathrm{in}$ & $210$ & $\mu$W& measured\\
			dn/dT & $2.35\times 10^{-4}$ & K$^{-1}$& ref. \cite{DellaCorte2000}$^,$ \cite{Skauli2003}\\
			$g_\mathrm{om}$ & $1.47\times 10^{20}$ & Hz m$^{-1}$& FEM\\
			$m_\mathrm{eff}$ & $255$ & pg& FEM\\
			$\omega_\mathrm{m}$ & $2\pi \times 131.7$ & MHz & measured\\
			$\Gamma_\mathrm{m}$ & $2\pi \times 135$ & kHz& measured\\
			$\alpha$ & $7.83$ & $\mu$N K$^{-1}$& FEM\\
			$\tau_\mathrm{th}$ & $3.99$ & $\mu$s& fit and FEM\\
			$R_\mathrm{th}$ & $5.64\times 10^{4}$ & K W$^{-1}$& fit and FEM\\
			$\beta_\mathrm{th}$ & $5.7\times 10^{-6}$ & K $^{-1}$& ref. \cite{Blakemore1982}\\
		\end{tabular}
	\end{ruledtabular}
\end{table}

To better appreciate the relative contributions of the photothermal force and radiation pressure in our experiments, in Fig.~\ref{fig:Figure3}~(c,d) we extract from our model the mechanical displacement modulation component $X$ as function of the modulation frequency (solid line) and compare it to situations where the photothermal forces only are switched-off (thin dashed line, $\alpha=0$) and where all thermal effects are switched off (thick dashed line, $\alpha$, R$_{th}=0$). At frequencies below $10^6$~Hz the amplitude of $X$ is three order of magnitude larger when the photothermal force is present (Fig.~\ref{fig:Figure3}~(c)). It reduces above the thermal frequency ($\sim 10^6$~Hz), following a first-order filter function. At the same time, a $\pi/2$ phase lag is present at modulation frequencies higher than $10^6$~Hz, which disappears when thermal effects are switched off. At even higher frequency, close to the mechanical resonance, the $X$ amplitude also increases by a decade when the photothermal force is present ($\alpha\neq0$), despite the two order of magnitudes difference between thermal and mechanical frequencies (Fig.~\ref{fig:Figure3}~(d)). A $\pi$ phase shift is retrieved in $X$ when scanning over the mechanical resonance, in accordance with an harmonic oscillator response.
This overall behavior is consistent with that of a damped mechanical oscillator driven by two forces, radiation pressure and photothermal, of different intensity and response function.

\begin{figure}[!h]
	\includegraphics[width=8.5cm]{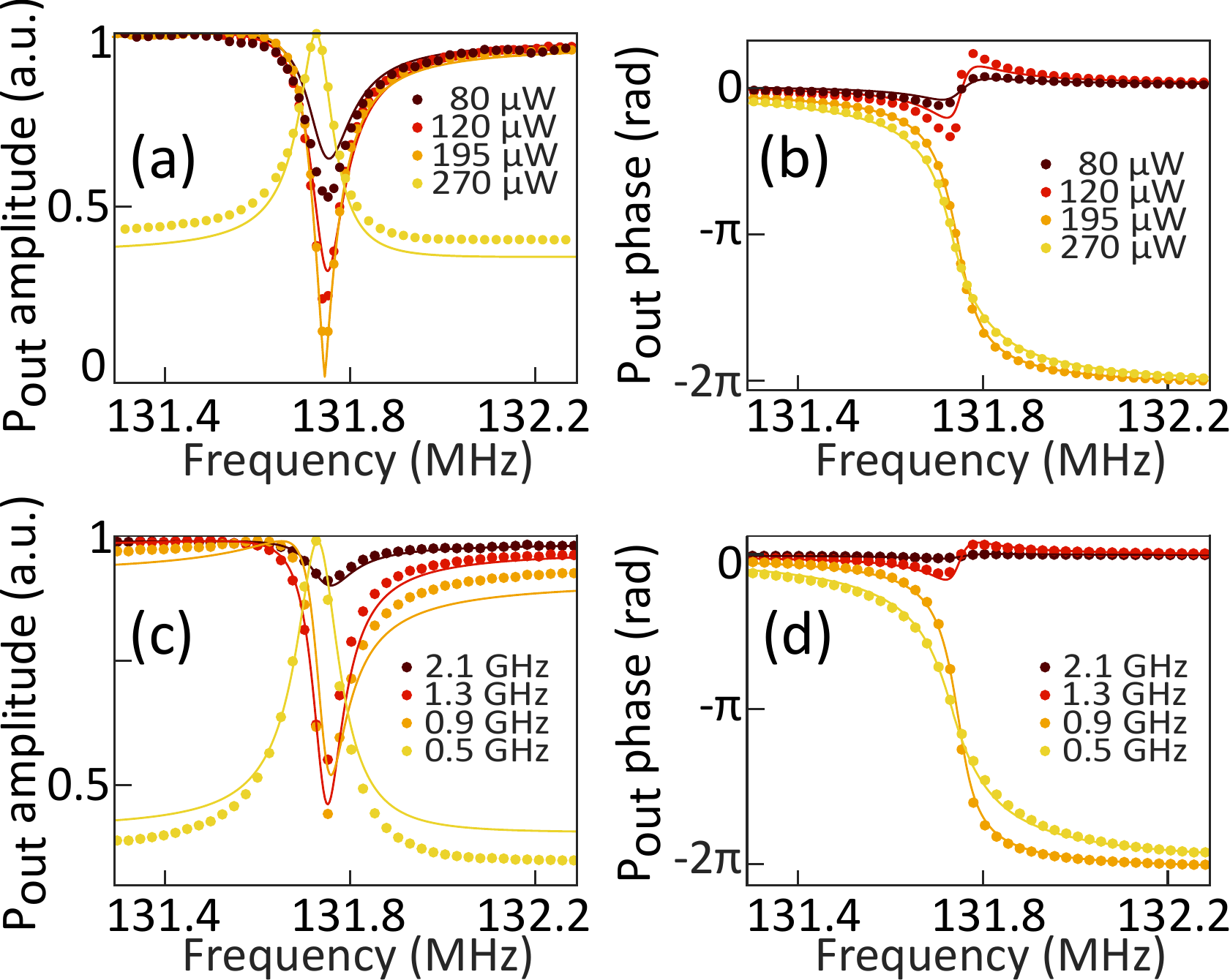}
	\caption{Amplitude and (unwrapped) phase response function of the demodulated optical signal when the modulation frequency is swept around the mechanical resonance frequency of RBM1 for different input powers (a,b) and laser-cavity detuning (c,d). Amplitude is normalized with respect to its maximum value over the frequency span. Dots represent experimental data while the model is shown as a solid line.}
	\label{fig:Figure4}
\end{figure}

In order to further test the validity of our model, which includes a photothermal force that has both linear and nonlinear components in the number of photons, we now systematically vary the incident optical power and the laser-cavity detuning.
Figure~\ref{fig:Figure4} reports the amplitude and phase response of the demodulated optical signal at the RBM1 resonance frequency, as a function of the power (a,b) and detuning (c,d). A red-shift of the mechanical resonance due to thermal softening of the material is noticeable when the number of intracavity photons increases, i.e. at larger power and/or smaller detuning. Here again, the full model (solid line) reproduces well the experimental data, all over the explored range. For the largest power and smaller detuning, TPA is twice as large as linear absorption. In this regime, it is responsible for increasing the cavity line width and gives rise to a dominating nonlinear photothermal force. For a given optical power and detuning, the performances of the sensor can also be optimized by adjusting the modulation depth ($\beta$). Such optimization is shown in the supplementary material, where measurements of the frequency stability of our optomechanical disk sensor reach down to $10^{-7}$.

\begin{figure}[!h]
	\includegraphics[width=8.5cm]{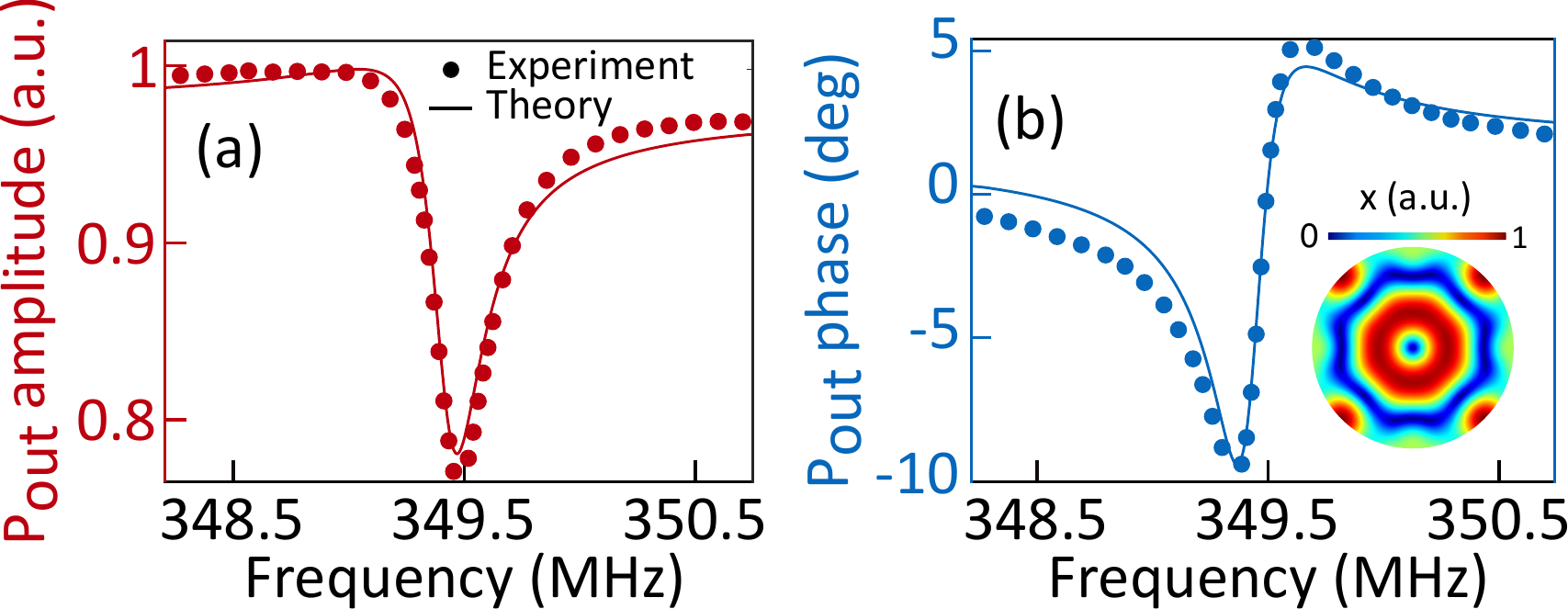}
	\caption{Amplitude (a) and phase (b) response function of the demodulated optical signal at RBM2 resonance frequency. The amplitude is normalized with respect to its maximum value over the frequency span. Dots represent experimental data, while the model is shown as a solid line. The RBM2 displacement profile is reported in inset.}
	\label{fig:Figure5}
\end{figure}

The model presented here can be applied to any mechanical mode of a sensor, by using the proper mechanical susceptibility, proper optomechanical coupling, and proper photothermal force. In Fig.~\ref{fig:Figure5} we report experimental data acquired on the second order RBM (RBM2), together with the fit by the model. Both the peak amplitude and phase jump are smaller with respect to those of RBM1. This is the consequence of a lower mechanical quality factor $ \left( Q_\mathrm{RBM2}/Q_\mathrm{RBM1}=0.4\right)$ and of an increased spectral distance to the thermal cut-off frequency, which reduces the photothermal actuation efficiency.\\

In conclusion, we have developed and systematically tested a model that correctly depicts all-optical actuation and detection of optomechanical devices operating at ultra-high frequency in the oscillating mode. In contrast to prior models, it does account for photothermal forces, both linear and nonlinear, while also embedding radiation pressure and electrostrictive effects, without limitation on the mechanical frequency range. These features are essential for a precise description of chip-based semiconductor optomechanical sensors working at room temperature, which are currently under development. Their modelling will enable the accurate interpretation of the demodulated sensor output when a physical signal (to be detected) triggers its response. 


\begin{acknowledgments}
The authors acknowledge support from the European Commission through the VIRUSCAN (731868) FET-open and NOMLI (770933) ERC projects, and from the Agence Nationale de la Recherche through the Olympia and QuaSert projects.
\end{acknowledgments}

\noindent The  data  that  support  the  findings  of  this  study  are  available from the corresponding authors upon reasonable request.
\bibliography{library}

\end{document}